\documentclass[11pt]{article}
\usepackage{amsfonts}
\usepackage{graphicx,color}
\usepackage{epstopdf}
\usepackage{epsfig}
\usepackage[breaklinks=true]{hyperref}
\usepackage{appendix}

\newcommand{\text}{\mathrm}

\newcommand{\beq}{\begin{equation}}
\newcommand{\eeq}{\end{equation}}
\newcommand{\bqa}{\begin{eqnarray}}
\newcommand{\eqa}{\end{eqnarray}}

\definecolor{green}{rgb}{0.00,0.50,0.00}

\begin{document}

\title{Quantum Trajectories for a Class of Continuous Matrix Product Input States}
\author{John E. Gough\thanks{Department of Mathematics and Physics, Aberystwyth University,
SY23 3BZ, Wales, United Kingdom. {\tt email: jug@aber.ac.uk}}, Matthew R. James\thanks{ARC Centre for Quantum Computation and Communication Technology, 
  Research School of Engineering, Australian National University, Canberra, ACT 0200, Australia. {\tt email: Matthew.James@anu.edu.au}
}, and Hendra I. Nurdin\thanks{ School of Electrical Engineering and Telecommunications, UNSW
Australia, Sydney, NSW 2052, Australia. {\tt email: h.nurdin@unsw.edu.au}}}
\date{\today}

\maketitle

\begin{abstract}
We introduce a new class of continuous matrix product (CMP) states and
establish the stochastic master equations (quantum filters) for an arbitrary
quantum system probed by a bosonic input field in this class of states. We
show that this class of CMP states arise naturally as outputs of a Markovian
model, and that input fields in these states lead to master and filtering
(quantum trajectory) equations which are matrix-valued. Furthermore, it is
shown that this class of continuous matrix product states include the
(continuous-mode) single photon and time-ordered multi-photon states.
\end{abstract}

\section{Introduction}

\begin{figure}[h]
\begin{center}
\includegraphics[scale=1.0]{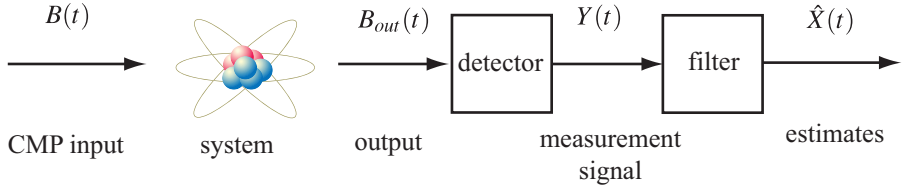} 
\end{center}
\caption{A schematic representation of a continuous measurement process,
where the measurement signal produced by a detector is filtered to produce
estimates $\hat X(t)=\protect\pi_t(X)=\mathrm{tr}[ \protect\rho(t) X] $ of
system operators $X$ at time $t$. The system is driven by a field in a CMP
state defined in this paper.}
\label{fig:system-filter}
\end{figure}

Continuous matrix product (CMP) states were introduced by Verstraete and
Cirac \cite{CMP_states}-\cite{HCOV13} as the generalization of finitely
correlated states to continuous-variable quantum input processes \cite
{GarZol00}. Here we introduce a new class of CMP states and derive the
quantum filtering (quantum trajectory) equations \cite{Q_Filter} for state
inputs in this class. We show that both the master equation and filter
equations become matrix-valued. In particular, we show that this class
includes (continuous-mode) single photon and multiphoton states of a boson
field. Whereas discrete matrix product states have been proposed to model
approximations an efficient simulation of certain continuous stochastic
master equation \cite{GM10}, our work here differs in so far as we wish to
deal with continuous variable models for an open quantum system with a
quantum input field in a state general enough to enable efficient derivation
of quantum trajectory equations (that is construct the quantum filter for
determining estimates of system operators) for important classes of
non-classical field states, Figure \ref{fig:system-filter}.

\begin{figure}[h]
\begin{center}
\includegraphics[scale=1.0]{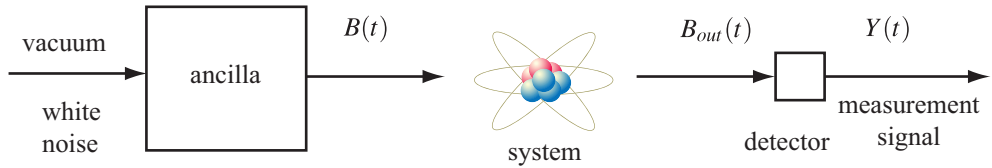} 
\end{center}
\caption{An ancilla system is used to model the effect of the CMP field
state for the input field $B(t)$ on the system.}
\label{fig:system-signal}
\end{figure}

As mentioned, a particularly important motivating class of inputs are
multi-photon states. 
The production and verification of single-photon states \cite
{single-photon-production} has become routine, achievable through a variety
of experimental architectures such as cavity quantum electrodynamics (QED),
quantum dots in semiconductors, and circuit QED. Single photon and
multi-photon states are important because they are of interest in various
applications; see, e.g., \cite{KLM01,RalGilMil03,GisRibTit02,CirZolKim97}
for single photon states and \cite{BCBC12,SZX13} for multi-photon states. We
shall first outline the solution in the standard case of Markovian systems
driven by a vacuum input field state in Section \ref{sec:quantum-io-models},
then give the generalization to our class of CMP state inputs. The filtering
equations make use of an ancilla cascaded with the system, as shown in
Figure \ref{fig:system-signal}.

Furthermore, it is shown that the class of continuous matrix product states
defined in this paper include the (continuous-mode) single photon and time-ordered 
multi-photon states, and we derive explicit Markovian generators for
these multi-photon states that allow the filtering equations for systems driven by
fields in these states to be obtained from the general formulas of
this paper.

The structure of this paper is as follows. Section \ref
{sec:quantum-io-models} provides a brief overview of quantum Markov
input-output models, quantum stochastic differential equations and the $%
(S,L,H)$ formalism, and the well-known quantum master and filtering
equations for open Markov models driven by a vacuum state input field. In
Section \ref{sec:CMP-state-intro} we introduce a new class of CMP states
that is defined in terms of a Hudson-Parthasarathy quantum stochastic
differential equation and which can be viewed as the output states of an $%
(S,L,H)$ model. Then in Section \ref{sec:filtering-CMP} we derive the
quantum master equation and quantum filtering (quantum trajectory) equations
for an open Markov model driven by a field in the newly defined CMP states,
in the form of matrix-valued equations with operator entries. This is
followed in Section \ref{sec:CMP-examples} by some explicit examples of the
novel CMP states: continuous-mode single photon states, and  time-ordered
continuous-mode multi-photon states. Section \ref{sec:conclusion} then
provides a summary of the contributions of this paper. The main text is also
supplemented by two appendices. Appendix \ref{app:two-photon-generator} details a
Markovian generator model for time-ordered two-photon states that is then generalized to
time-ordered multi-photon states in Appendix \ref{app:n-photon-generator}.

\section{Quantum Input-Output Models}
\label{sec:quantum-io-models}
We review quantum Markov input-output models.
For simplicity, we shall consider single-input single-output (SISO) models
only, however the ideas are readily extended to multiple inputs. To describe
external inputs, we fix a Fock space $\mathfrak{F}$ for the quantum field
inputs. Formally, we have quantum input processes $b_{\mathrm{in}}( t) $
satisfying a singular canonical commutation relations (CCR) $\left[ b_{%
\mathrm{in}}(t),b_{\mathrm{in}}(s)^{\ast }\right] =\delta ( t-s) $, and
define regular processes of annihilation, creation and number: $B_{\mathrm{in%
}}( t) =\int_{0}^{t}b_{\mathrm{in}}(s)ds$, $B_{\mathrm{in}}( t) ^{\ast
}=\int_{0}^{t}b_{\mathrm{in}}(s)^{\ast }ds$ and $\Lambda _{\mathrm{in}}( t)
=\int_{0}^{t}b_{\mathrm{in}}(s)^{\ast }b_{\mathrm{in}}(s)ds$. We shall use
the framework of the Hudson-Parthasarathy calculus of quantum stochastic
integration with respect to these processes \cite{HudPar84}, which contains
the quantum input theory of Gardiner \cite{GarZol00} as a special case. The
Fock vacuum state will be denoted by $|\Omega \rangle $.

We shall be interested in a new class \emph{continuous matrix product (CMP)
states} which are of the form 
\begin{equation}
\left| \Psi ( T) \right \rangle \equiv \vec{T}e^{\int_{0}^{T}R( t) \otimes
dB_{\mathrm{in}}( t) ^{\ast }+Q( t) \otimes dt}\,\left| \phi \right\rangle
\otimes \left| \Omega \right\rangle  \label{eq:cmp_state}
\end{equation}
with $T>0$, $R( \cdot ) $ and $Q( \cdot ) $ $\mathbb{C}^{D\times D}$-valued
functions, and $\phi $ a\ fixed unit vector in $\mathbb{C}^{D}$. We may think
of $\mathbb{C}^{D}$ as being an auxiliary finite-dimensional Hilbert space.
Without loss of generality one takes $Q( t) \equiv -\frac{1}{2}R( t) ^{\ast
}R( t) -iH_{\mathrm{aux}}( t) $, where $H_{\mathrm{aux}}$ is Hermitean
valued.

Note that this class of CMP states is distinct from the one introduced by
Verstraete-Cirac \cite{CMP_states} which defines an unnormalized CMP state
as a state of the form 
\[
\left| \Psi ( T)\rangle \right.\rangle_{\mathrm{VC}} \equiv \mathrm{tr}_{%
\mathrm{aux}}\left( B \vec{T}e^{\int_{0}^{T}R( t) \otimes dB_{\mathrm{in}}(
t) ^{\ast }+Q( t) \otimes dt}\right)\,\left| \Omega \right. \rangle , 
\]
where $B$ is a fixed operator on the ancilla space called the boundary
operator. Typical choices studied are $B =I$ and $B= \vert e_k \rangle
\langle e_j \vert$, where $\left \{ \vert e_k \rangle \right\}$ is an
orthonormal basis for the ancilla space $\mathbb{C}^D$. Whereas $\left| \Psi (
T)\rangle \right.\rangle_{\mathrm{VC}} $ is a pure state on the quantum
field Hilbert space, our new CMP state (\ref{eq:cmp_state}) is a pure state
on the composite auxiliary and quantum field Hilbert space. Nonetheless, our
CMP states share the continuous product property of the Verstraete-Cirac CMP
states.

To connect the classes, we introduce a Fock space vectors $| \Psi_{jk} (t)
\rangle $ defined by 
\[
\langle \Phi \vert \Psi_{jk} ( T)\rangle := \langle e_j \otimes \Phi | \vec{T%
}e^{\int_{0}^{T} R( t) \otimes dB_{\mathrm{in}}( t)^\ast +Q( t) \otimes dt}
\, | e_k \otimes \Omega \rangle , 
\]
for arbitrary Fock space vector $| \Phi \rangle $. It follows that we work
with the $\mathbb{C}^D$-valued Fock space vector 
\[
\left| \Psi ( T) \right \rangle = \left[ \sum_k \left| \Psi_{jk} ( T)
\right
\rangle \, \langle e_k \vert \phi \rangle \right]_{j=1}^D , 
\]
while Verstraete-Cirac work with Fock space vectors 
\[
\left| \Psi ( T) \right \rangle_{VC} = \sum_{jk} B_{kj} \, \left| \Psi_{jk}
( T) \right \rangle . 
\]

We remark that the treatment of Cirac, Verstraete, et al., employs a spatial
observable $x$ rather than a time variable. In our interpretation we think
of a travelling input field, which we may take as a quantum optical field
propagating at the speed of light $c$, so that initial the element of the
field that will interaction in a Markov manner with the system at time $t$
will have to travel the distance $x=ct$ to the system. In fact, Haegeman,
Cirac and Osborne have noted that the spatial interpretation for their CMP
states is equivalent to this type of Markov model, see the discussion at the
end of section E of \cite{HCOV13} on physical interpretations. Indeed, they
note that $\vert \Psi_{jk} (T) \rangle $, corresponding to their choice $B =
\vert e_k \rangle \langle e_j \vert$, corresponds to an initialization of
the ancilla in state $\vert e_j \rangle$ followed by a post-selection of
state $\vert e_k \rangle$ by measurement at time $T$.

In either case, the natural mathematical setting for the theory is the
Hudson-Parthasarathy quantum stochastic calculus \cite{HudPar84}. Let us fix
a Hilbert space $\mathfrak{h}_{0}$ which we call the initial space. On the
joint Hilbert space $\mathfrak{h}_{0}\otimes \mathfrak{F}$ we may consider
the quantum stochastic differential equation (QSDE) 
\begin{eqnarray}
V( t,s) =I+\int_{s}^{t}\{( S-I_{\mathrm{sys}}) \otimes d\Lambda_{\mathrm{in}%
} (\tau )+L\otimes dB_{\mathrm{in}}(\tau )^{\ast }  \nonumber \\
-L^{\ast }S\otimes dB_{\mathrm{in}}(\tau )-( \frac{1}{2}L^{\ast }L+iH)
\otimes d\tau \}V( \tau ,s)
\end{eqnarray}
for $t>s$. We then have the propagation law $V( t,s) =V( t,r) V( r,s) $
whenever $t>r>s$. The differentials are all understood to be future-pointing
(It\={o} convention) and by the CCR we have the following non zero products 
\begin{eqnarray*}
dB_{\mathrm{in}}( t) dB_{\mathrm{in}}( t) ^{\ast } &=&dt,\,\quad d\Lambda _{%
\mathrm{in}}( t) dB_{\mathrm{in}}( t) ^{\ast }=dB_{\mathrm{in}}( t) ^{\ast }
\\
dB_{\mathrm{in}}( t) d\Lambda _{\mathrm{in}}( t) &=&dB_{\mathrm{in}}( t)
\quad d\Lambda _{\mathrm{in}}( t) d\Lambda _{\mathrm{in}}( t) =d\Lambda _{%
\mathrm{in}}( t) .
\end{eqnarray*}
Provided the operator $S$ is unitary, and $H$ is self-adjoint, there will
exist a unique solution to the QSDE which is unitary. We shall refer to the
triple $( S,L,H) $ as the (possibly time-dependent) \emph{%
Hudson-Parthasarathy} (HP) parameters. In particular, we shall be interested
in a fixed origin of time and will set $V( t) =V( t,0) $.

We now fix $\mathfrak{h}_0$ as the Hilbert space $\mathfrak{h}_{\mathrm{sys}%
} $ of a quantum system of interest. For a given system observable $X$, we
set $j_{t}(X)=V( t) ^{\ast }( X\otimes I_{\mathfrak{F}}) V( t) $ in which
case we obtain the QSDE 
\begin{eqnarray}
dj_{t}( X) &=&j_{t}( \mathcal{L}_{00}^{\mathrm{sys}}X) dt+j_{t}( \mathcal{L}%
_{10}^{\mathrm{sys}}X) dB_{\mathrm{in}}(t)^{\ast }  \nonumber \\
&&+j_{t}( \mathcal{L}_{01}^{\mathrm{sys}}X) dB_{\mathrm{in}}(t)+j_{t}(%
\mathcal{L}_{11}^{\mathrm{sys}}X)d\Lambda _{\mathrm{in}}(t)
\label{eq:Heisenberg}
\end{eqnarray}
where we have the following superoperators 
\begin{eqnarray}
\mathcal{L}_{00}^{\mathrm{sys}}X &=&\frac{1}{2}\left[ L^{\ast },X\right] L+%
\frac{1}{2}L^{\ast }\left[ X,L\right] -i\left[ X,H\right]  \nonumber \\
\mathcal{L}_{10}^{\mathrm{sys}}X &=&S^{\ast }\left[ X,L\right] ,\mathcal{L}%
_{01}^{\mathrm{sys}}X=\left[ L^{\ast },X\right] S  \nonumber \\
\mathcal{L}_{11}^{\mathrm{sys}}X &=&S^{\ast }XS-X
\label{eq:Evans_Hudson_Maps}
\end{eqnarray}
known as the Evans-Hudson maps (note that $\mathcal{L}_{11}^{\mathrm{sys}}$
is a Lindbladian). The output fields are given by 
\begin{equation}
B_{\mathrm{out}}( t) =V( t) ^{\ast }( I_{\mathrm{sys}}\otimes B_{\mathrm{in}%
}( t) ) V( t) ,  \label{eq:output}
\end{equation}
etc., and we find from the quantum It\={o} calculus $dB_{\mathrm{out}}( t)
=j_{t}( S) dB_{\mathrm{in}}( t) +j_{t}(L) dt$. We remark that we may also
write $B_{\mathrm{out}}( t) =V( T) ^{\ast }( I_{\mathrm{sys}}\otimes B_{%
\mathrm{in}}( t) ) V( T) $ whenever $T>t$ since we have $V( T) =V( T,t) V(
t) $ and the unitary $V( T,t) $ acts non-trivially only on the component of
the Fock space generated by fields on the time interval $\left[ t,T\right] $
and in particular commutes with $I_{\mathrm{sys}}\otimes B_{\mathrm{in}}( t) 
$, and so can be removed.

Let us fix a state $\eta \in \mathfrak{h}_{\mathrm{sys}}$, then for $t\geq 0$
we may define the expectation $\mathbb{E}_{\eta \Omega }\left[ \cdot \right]
=\langle \eta \otimes \Omega |\cdot |\eta \otimes \Omega \rangle $ and for $%
X $ a system observable we define $\mathbb{E} _{t}^{\mathrm{vac}}( \cdot )$ by 
\[
\mathbb{E} _{t}^{\mathrm{vac}}( X) :=\mathbb{E}_{\eta \Omega }\left[ j_{t}( X) %
\right]\equiv tr_{\mathrm{sys}}\left\{ \varrho _{\mathrm{sys}}X\right\} , 
\]
which introduces the density matrix $\varrho _{\mathrm{sys}}$. We obtain the 
\textit{Ehrenfest equation } $\frac{d}{dt}\mathbb{E} _{t}^{\mathrm{vac}}( X) =%
\mathbb{E} _{t}^{\mathrm{vac}}( \mathcal{L}_{00}^{\mathrm{sys}}X) $, with
equivalent master equation $\frac{d}{dt}\varrho _{\mathrm{sys}}=\mathcal{L}%
_{00}^{\mathrm{sys}\star }\varrho _{\mathrm{sys}}$ where $\mathcal{L}_{00}^{%
\mathrm{sys}\star }\rho =L\rho L^{\ast }-\frac{1}{2}\left\{ \rho ,L^{\ast
}L\right\} _{+}+i\left[ \rho ,H\right] $.


We now consider the following continuous measurements of the output field 
\cite{Q_Filter}

\begin{enumerate}
\item  Homodyne $Z( t) =B_{\mathrm{in}}( t) +B_{\mathrm{in}}^{\ast }( t) ;$

\item  Number counting $Z( t) = \Lambda_{\mathrm{in}}( t) .$
\end{enumerate}

In both cases, the family $\left\{ Z( t) :t\geq 0\right\} $ is
self-commuting, as is the set of observables 
\begin{equation}
Y( t) =V( t) ^{\ast }( I_{\mathrm{sys}}\otimes Z( t) ) V( t)
\end{equation}
which constitute the actual measured process. We note the non-demolition
property $\left[ j_{t}(X),Y_{s}\right] =0$ for all $t\geq s$. The aim of
filtering theory is to obtain a tractable expression for the least-squares
estimate of $j_{t}( X) $ given the output observations $Y( \cdot ) $ up to
time $t$. Mathematically, this is the conditional expectation 
\begin{equation}
\pi _{t}( X) =\mathbb{E}_{\eta \Omega }\left[ j_{t}( X) |\mathcal{Y}_{t}\right]
\end{equation}
onto the measurement algebra $\mathcal{Y}_{t}$ generated by $Y( s) $ for $%
s\leq t$.

The filters are given respectively by \cite{Q_Filter} \newline
\textbf{Homodyne: } 
\begin{eqnarray*}
d\pi _{t}(X) &=&\pi _{t}(\mathcal{L}_{00}X)dt \\
&&+\left\{ \pi _{t}(XL+L^{\ast }X)-\pi _{t}(X)\pi _{t}(L+L^{\ast })\right\}
[dY(t)-\pi _{t}(L+L^{\ast })dt],
\end{eqnarray*}
\textbf{Number counting: } 
\begin{eqnarray*}
d\pi _{t}(X) &=&\pi _{t}(\mathcal{L}_{00}X)dt \\
&&+\left\{ \pi _{t}(L^{\ast }XL)/\pi _{t}(L^{\ast }L)-\pi _{t}(X)\right\}
[dY(t)-\pi _{t}(L^{\ast }L)dt].
\end{eqnarray*}

Note that there can be other types of measurements performed since a quantum
master equation can be ``unravelled'' in infinitely many ways \cite
{Carm93,WM10}, leading to infinitely many possible stochastic master
equations. However, the two types of measurements that we consider here are
the most natural and commonly performed in the laboratory.

\section{CMP States}
\label{sec:CMP-state-intro} CMP states may likewise be considered as $(
S,L,H) $-model output states. Here we take the initial Hilbert space to be
the auxiliary space $\mathfrak{h}_{\mathrm{aux}}=\mathbb{C}^{D}$. The HP
parameters are taken to be $( I_{\mathrm{aux}},R( \cdot ) ,H_{\mathrm{aux}}(
\cdot ) ) $, which may be time-dependent. The state (\ref{eq:cmp_state}) is
then realized as 
\begin{equation}
\left| \Psi ( T) \right\rangle =V_{\mathrm{aux}}( T) \,\left| \phi
\right\rangle \otimes \left| \Omega \right\rangle  \label{eq:CMP_state_II}
\end{equation}
where $V_{\mathrm{aux}}$ is the associated unitary.

\bigskip

\textbf{Definition:} \textit{\ Suppose that for a fixed auxiliary space $%
\mathfrak{h}_{\mathrm{aux}}$, a prescribed set of HP parameters, and a fixed
unit vector $\phi \in \mathfrak{h}_{\mathrm{aux}}$ as above, the vector
states $\Psi ( T) $ have a well-defined limit in norm $\Psi ( \infty ) \in %
\mathfrak{h}_{\mathrm{aux}}\otimes \mathfrak{F}$, as $T \to \infty$. Then we
refer to $\Psi ( \infty ) $ as an asymptotic continuous matrix product state.%
}

For arbitrary operator $A$ on the auxiliary space, we have the expectation 
\[
\langle \phi \otimes \Omega |V_{\mathrm{aux}}( t) ^{\ast }( A\otimes I_{%
\mathfrak{F}}) V_{\mathrm{aux}}( t) |\phi \otimes \Omega \rangle \equiv tr_{%
\mathrm{aux}}\left\{ \varrho _{\mathrm{aux}}( t) A\right\} 
\]
which defines the reduced density matrix $\varrho _{\mathrm{aux}}( t) $. We
note that it satisfies the master equation $\frac{d}{dt} \varrho _{\mathrm{%
aux}}=\mathcal{L}_{00}^{\mathrm{aux}\star }\varrho _{\mathrm{aux}}$ with $%
\varrho _{\mathrm{aux}}( 0) =|\phi \rangle \langle \phi |$.

\section{Filtering for CMP State Inputs}
\label{sec:filtering-CMP} We now wish to consider a system with Hilbert
space $\mathfrak{h}_{\mathrm{sys}}$\ and HP parameters $( S,L,H) $ driven by
an input field in a CMP state $|\Psi _{\infty }\rangle $ on noise space $%
\mathfrak{h}_{\mathrm{aux}}\otimes \mathfrak{F}$. To this end we have the
expectation 
\[
\mathbb{E} _{t}^{\mathrm{cmp}}( X) =\mathbb{E}_{\eta \Psi _{\infty }}\left[ j_{t}(
X) \otimes I_{\mathrm{aux}}\right] . 
\]
Our aim is to derive the master and filtering equations based on the CMP
state.

\subsection{Cascade Realization}
As we are effectively working on the joint Hilbert space $\mathfrak{h}_{%
\mathrm{sys}}\otimes \mathfrak{h}_{\mathrm{aux}}\otimes \mathfrak{F}$, it is
convenient to take the initial space to be $\mathfrak{h}_{0}= \mathfrak{h}_{%
\mathrm{sys}}\otimes \mathfrak{h}_{\mathrm{aux}}$. With respect to this
decomposition we introduce the pair of HP parameters 
\begin{eqnarray*}
&&G_{\mathrm{sys}}=( S\otimes I_{\mathrm{aux}},L\otimes I_{\mathrm{aux}%
},H\otimes I_{\mathrm{aux}}) \\
&&G_{\mathrm{aux}}=( I_{\mathrm{sys}}\otimes I_{\mathrm{aux}},I_{\mathrm{sys}%
}\otimes R( t) ,I_{\mathrm{sys}}\otimes H_{\mathrm{aux}}( t) )
\end{eqnarray*}
and denote the associated unitary processes by $\tilde{V}_{\mathrm{sys}}( t) 
$ and $\tilde{V}_{\mathrm{aux}}( t) $ respectively. We then have 
\begin{eqnarray}
\mathbb{E} _{t}^{\mathrm{cmp}}( X) &=& \lim_{T\to \infty} \mathbb{E}_{\eta \Psi (
T) }\left[ \tilde{V}_{\mathrm{sys}}( t) ^{\ast }( X\otimes I_{\mathrm{aux}%
}\otimes I_{\mathfrak{F}}) \tilde{V}_{\mathrm{sys}}( t) \right]  \nonumber \\
&=&\mathbb{E}_{\eta \phi \Omega }\left[ \tilde{V}( t) ^{\ast }( X\otimes I_{%
\mathrm{aux}}\otimes I_{\mathfrak{F}}) \tilde{V}( t) \right]
\label{eq:switch}
\end{eqnarray}
where $\tilde{V}( t) =\tilde{V}_{\mathrm{sys}}( t) \tilde{V}_{\mathrm{aux}}(
t) $, and we use the fact that $\tilde{V}_{\mathrm{sys}} (t) \tilde{V}_{%
\mathrm{aux}}(T) =\tilde{V}_{\mathrm{aux}}(T,t) \tilde{V} (t)$. From the
quantum It\={o} calculus it is easy to see that $\tilde{V}$ is associated to
the HP parameters (this is a special case of the series product for cascaded
network consisting of the auxiliary model fed into the system model \cite{JG}%
) 
\begin{eqnarray}
\tilde{S} &=&S\otimes I_{\mathrm{aux}},\quad \tilde{L}( t) =L\otimes I_{%
\mathrm{aux}}+S\otimes R( t) ,  \nonumber \\
\tilde{H}( t) &=&H\otimes I_{\mathrm{aux}}+I_{\mathrm{sys}}\otimes H_{%
\mathrm{aux}}(t)  \nonumber \\
&&+\frac{1}{2i}L^{\ast }S\otimes R( t) -\frac{1}{2i}S^{\ast }L\otimes R( t)
^{\ast }.  \label{eq:SLH_cascade}
\end{eqnarray}

Let us introduce the more general expectation 
\[
\tilde{\mathbb{E}}_{t}(X\otimes A)=\mathbb{E}_{\eta \phi \Omega }\left[ \tilde{V}%
(t)^{\ast }(X\otimes A\otimes I_{\mathfrak{F}})\tilde{V}(t)\right] 
\]
for $X$ and $A$ system and auxiliary operators respectively. We then have $%
\tilde{\mathbb{E}}_{t}(X\otimes I_{\mathrm{aux}})=\mathbb{E}_{t}^{\mathrm{cmp}}(X)$%
. Now 
\[
\frac{d}{dt}\tilde{\mathbb{E}}_{t}(X\otimes A)=\tilde{\mathbb{E}}_{t}(\tilde{%
\mathcal{L}}_{00}(X\otimes A)) 
\]
where $\tilde{\mathcal{L}}_{00}$ is the Lindbladian corresponding to the HP
parameters $(\tilde{S},\tilde{L},\tilde{H})$. We now note the following
identity 
\begin{eqnarray}
\tilde{\mathcal{L}}_{00}(X\otimes A) &=&\mathcal{L}_{00}^{\mathrm{sys}%
}X\otimes A  \nonumber \\
&&+\mathcal{L}_{01}^{\mathrm{sys}}X\otimes AR+\mathcal{L}_{10}^{\mathrm{sys}%
}X\otimes R^{\ast }A+\mathcal{L}_{11}^{\mathrm{sys}}X\otimes R^{\ast }AR 
\nonumber \\
&&+X\otimes \mathcal{L}_{00}^{\mathrm{aux}}A.  \label{eq:Lindblad_identity}
\end{eqnarray}

\subsection{Matrix Form of the Master and Filter Equations}
Let us fix an orthonormal basis $\left\{ e_{n}\right\} $ for the auxiliary
space $\mathbb{C}^{D}$. We introduce the $D\times D$ matrix $\Upsilon _{t}( X) $
with entries 
\[
\Upsilon _{t}^{nm}( X) =\tilde{\mathbb{E}}_{t}( X\otimes E_{mn}) 
\]
where $E_{mn}=|e_{m}\rangle \langle e_{n}|$. The CMP expectation is then 
\[
\mathbb{E} _{t}^{\mathrm{cmp}}( X) =tr\Upsilon _{t}( X) , 
\]
and more generally $\tilde{\mathbb{E}}_{t}( X\otimes A) =tr\left\{ \Upsilon
_{t}( X) A\right\} $. Unlike the vacuum case, there is no single closed
master equation for $\mathbb{E} _{t}^{\mathrm{cmp}}( X) $, and instead we must
solve a system of matrix equations.

\subsubsection{CMP Master Equation:}
This takes the form
\begin{eqnarray}
\frac{d}{dt}\Upsilon _{t}( X) &=&\Upsilon _{t}( \mathcal{L}_{00}^{\mathrm{sys%
}}X) +R\Upsilon _{t}( \mathcal{L}_{01}^{\mathrm{sys}}X) +\Upsilon _{t}( 
\mathcal{L}_{10}^{\mathrm{sys}}X) R^{\ast}  \nonumber \\
&&+ \mathcal{L}_{00}^{\mathrm{aux} \ast} \Upsilon_t(X) +R\Upsilon _{t}( 
\mathcal{L}_{11}^{\mathrm{sys}}X) R^{\ast },  \label{eq:Matrix_Master}
\end{eqnarray}
where 
\begin{equation}
\mathcal{L}_{00}^{\mathrm{aux} \ast} (A) = \frac{1}{2}[ L A, L^\ast] + \frac{%
1}{2}[ L, A L^\ast].
\end{equation}
These are obtained by extending the standard master equation to include the
auxiliary system and using (\ref{eq:Lindblad_identity}).

\subsubsection{CMP Filter Equation:}
Similarly we can consider the total filter 
\[
\tilde{\Pi _{t} }( X\otimes A ) =\mathbb{E}_{\eta \Psi _{\infty }}\left[ \tilde{%
V}_{\mathrm{sys}}( t) ^{\ast }( X\otimes A \otimes I_{\mathfrak{F}}) \tilde{V%
}_{\mathrm{sys}}( t) |\mathcal{Y}_{t} \right] , 
\]
and introduce the $D\times D$ matrix $\Pi _{t}( X) $ with entries 
\[
\Pi _{t}^{nm}( X) =\tilde{\Pi _{t} }( X\otimes E_{mn} ). 
\]
The CMP filter is then $\Pi^{\mathrm{cmp}} _{t}( X) =\tilde{\Pi _{t} }(
X\otimes I ) \equiv \mathrm{tr}_{\mathrm{aux}} \{ \Pi_t (X)\}$, where $%
\mathrm{tr}_{\mathrm{aux}}(\cdot)$ denotes the partial trace over the
auxiliary Hilbert space. We can again determine the explicit form of the
filter equations for both homodyne and number counting cases. We may take
the form of the previous filter equations and extend as above. The resulting
equations are as follows:


\subsubsection{Homodyne CMP State Filter}
The filter is
\begin{eqnarray}
d\Pi _{t}( X) &=&\{\Pi _{t} ( \mathcal{L}_{00}^{\mathrm{sys}}X) +R\Pi _{t}( 
\mathcal{L}_{01}^{\mathrm{sys}}X) +\Pi _{t}( \mathcal{L}_{10}^{\mathrm{sys}%
}X) R^{\ast }+R\Pi _{t}( \mathcal{L}_{11}^{\mathrm{sys}}X) R^{\ast } 
\nonumber \\
&& + \mathcal{L}_{00}^{\mathrm{aux}\star }\Pi_t (X) \}dt  \nonumber \\
&&+\{\Pi _{t}( XL+L^{\ast }X) +R\Pi _{t}( XS) +\Pi _{t}( S^{\ast }X) R^{\ast
}-\lambda _{t}\,\Pi _{t}( X) \}  \nonumber \\
&& \times \left[ dY-\lambda _{t}dt\right] ,  \label{eq:Homodyne_CMP_filter}
\end{eqnarray}
where $\mathcal{L}_{00}^{\mathrm{aux}\star }\Pi =R\Pi R^{\ast }-\frac{1}{2}%
\left\{ \Pi ,R^{\ast }R\right\} _{+}+i\left[ \Pi ,H_{\mathrm{aux}}\right] $
and 
\[
\lambda _{t}=tr \left\{ \Pi _{t}( L+L^{\ast }) +R\Pi _{t}( S) +\Pi _{t}(
S^{\ast }) R^{\ast } \right\}. 
\]

\subsubsection{Number Counting CMP Filter}
The filter is
\begin{eqnarray}
d\Pi _{t}( X) &=&\{\Pi _{t}( \mathcal{L}_{00}^{\mathrm{sys}}X) +R\Pi _{t}( 
\mathcal{L}_{01}^{\mathrm{sys}}X) +\Pi _{t}( \mathcal{L}_{10}^{\mathrm{sys}%
}X) R^{\ast }+R\Pi _{t}( \mathcal{L}_{11}^{\mathrm{sys}}X) R^{\ast } 
\nonumber \\
&& + \mathcal{L}_{00}^{\mathrm{aux}\star }\Pi_t (X) \}dt  \nonumber \\
&&+\nu _{t}^{-1}\{\Pi _{t}( L^{\ast }XL) +R\Pi _{t}( L^{\ast }XS) +\Pi _{t}(
S^{\ast }XL) R^{\ast }  \nonumber \\
&& +R\Pi _{t}( S^{\ast }XS) R^{\ast }-\nu _{t}\Pi _{t}( X) \} \left[ dY-\nu
_{t}dt\right] ,  \label{eq:number_CMP_filter}
\end{eqnarray}
where $\nu _{t}=tr\left\{ \Pi _{t}( L^{\ast }L) +R\Pi _{t}( L^{\ast }S) +\Pi
_{t}( S^{\ast }L) R^{\ast }+RR^{\ast }\right\} $.

\bigskip

\subsubsection{Stochastic Master Equation}

We may introduce a density matrix $\varrho _{t}$ over the system and
auxiliary space such that $\Pi _{t}\left( X\right) \equiv tr\left\{ \varrho
_{t}X\right\} $. This may be viewed as again as a $D\times D $ matrix whose
entries are trace class operators on the system space, and $tr\left\{
\varrho_{t} X\right\}$ denotes taking the trace on the $D\times D$ matrix $%
\varrho_{t} X =[\varrho_{t}^{nm} X]_{n,m=1,2,\ldots,D}$. The corresponding
stochastic master equation for $\varrho _{t}$ is then 
\[
d\varrho _{t}=\mathcal{L}^\star \varrho _{t}dt+\left\{ 
\begin{array}{cc}
\left[ \tilde{L}\varrho _{t}+\varrho _{t}\tilde{L}^{\ast }-\lambda
_{t}\varrho _{t}\right] \left[ dY\left( t\right) -\lambda _{t}dt\right] , & 
\mathrm{homodyne;} \\ 
\frac{1}{\nu _{t}}\left[ \tilde{L}\varrho _{t}\tilde{L}_{t}-\nu _{t}\varrho
_{t}\right] \left[ dY\left( t\right) -\nu _{t}dt\right] , & \mathrm{number \
counting};
\end{array}
\right. 
\]
where the dynamical term is 
\begin{eqnarray*}
\mathcal{L}^\star \varrho &=& \left( \mathcal{L}_{00}^{\mathrm{sys}\star
}\otimes I_{\mathrm{aux}}\right) \left( \varrho \right) +\left( \mathcal{L}%
_{01}^{\mathrm{sys}\star }\otimes I_{\mathrm{aux}}\right) \left( \varrho
R\right) +\left( \mathcal{L}_{10}^{\mathrm{sys}\star }\otimes I_{\mathrm{aux}%
}\right) \left( R^{\ast }\varrho \right)  \nonumber \\
&& +\left( \mathcal{L}_{11}^{\mathrm{sys}\star }\otimes I_{\mathrm{aux}%
}\right) \left( R^{\ast }\varrho R\right) +\left( I_{\mathrm{sys}}\otimes 
\mathcal{L}_{00}^{\mathrm{aux}\star }\right) \left( \varrho \right)
\end{eqnarray*}
and $\lambda _{t}=tr\left\{ \varrho _{t}(\tilde{L}+\tilde{L}^{\ast
})\right\} $, $\nu _{t}=tr\left\{ \varrho _{t}\tilde{L}^{\ast }\tilde{L}%
\right\} $. Here we use the usual convention of $\mathcal{J}^\star$ for the
dual of a superoperator $\mathcal{J}$, and that $\mathcal{J} \otimes I_{%
\mathrm{aux}}$ acting on a matrix with operator entries $\varrho^{nm}$
yields the matrix with entries $\mathcal{J} ( \varrho^{nm} )$. 

\section{Examples}

\label{sec:CMP-examples}

\subsection{Single Photon Sources}

As a special example of an asymptotic CMP state, let us take $D=2$ and fix $%
R( t) =\frac{1}{\sqrt{w( t) }}\xi ( t) \sigma _{-},\quad H_{\mathrm{aux}}=0$%
, where $\xi $ is a normalized square-integrable function on $[0,\infty )$, $%
w( t) =\int_{t}^{\infty }\left| \xi ( s) \right| ^{2}ds $, and $\sigma _{-}$
is the lowering operator from the upper state $|\uparrow \rangle $ to the
ground state $|\downarrow \rangle $ on $\mathfrak{h}_{\mathrm{aux}}=\mathbb{C}%
^{2}$. We take the initial state to be $|\phi \rangle =|\uparrow \rangle $.
\ The interpretation is that we have a two-level atom prepared in its
excited state $|\uparrow \rangle $ and coupled to the vacuum input field. At
some stage the atom decays through spontaneous emission into its ground
state $|\downarrow \rangle $ creating a single photon in the output. The
Schr\"{o}dinger equation for $|\Psi _{t}\rangle =V_{\mathrm{aux}%
}(t)|\uparrow \rangle \otimes |\Omega \rangle $ is $d|\Psi _{t}\rangle =%
\left[ \lambda ( t) \sigma _{-}dB_{t}^{\ast }-\frac{1}{2}\left| \lambda ( t)
\right| ^{2}\sigma _{+}\sigma _{-}dt\right] \,|\Psi _{t}\rangle $, where $%
\lambda ( t) =\frac{1}{\sqrt{w( t) }}\xi ( t) $ and it is easy to see that
this has the exact solution $|\Psi _{T}\rangle =\sqrt{w( T) }|\uparrow
\rangle \otimes |0\rangle +|\downarrow \rangle \otimes B_{\mathrm{in,}%
T}^{\ast }(\xi )|\Omega \rangle $where $B_{\mathrm{in,}T}^{\ast }( \xi )
=\int_{0}^{T}\xi _{t}dB_{\mathrm{in}}^{\ast }( t) $. As $w( \infty )
=\left\| \xi \right\| ^{2}=1$, we therefore generate the limit state $|\Psi
_{\infty }\rangle =|\downarrow \rangle \otimes B_{\mathrm{in}}^{\ast }( \xi
) |\Omega \rangle $. In this way we engineered a single photon with
one-particle state $\vert 1_\xi \rangle = B^\ast (\xi) \vert \Omega \rangle $%
. In the single-photon case, we encounter $2\times 2$ systems of equations.
As $R(t)^2 = 0$ we have some hierarchical simplification in these systems.
The matrix master equation was effectively derived by Gheri et al., \cite
{GheEllPelZol99} in 1998, however the filtering equations only more recently
in \cite{GJN13,GJNC12}.

\subsection{Time-Ordered Multi-Photon Sources}

Recently, quantum filters for multiple photon input states have been derived 
\cite{SZX13}, extending the work mentioned in the previous section. The
derivation of the multi-photon filters in \cite{SZX13} employed a
non-Markovian embedding technique, generalizing the approach of \cite{GJN13}. 
Here we indicate briefly that the filter equation for time-ordered multi-photon inputs may be
derived using the CMP approach presented in the present paper, generalizing
the Markovian embedding approach of \cite{GJNC12} for the single photon case.

Consider the $n$-photon state 
\begin{eqnarray*}
|\overrightarrow{\xi _{n},\cdots ,\xi _{1}}\rangle &=&\frac{1}{\prod_{k=1}^{n-1} \sqrt{\int_{0}^{\infty} |\xi_{k}(s)|^2
w_{k+1}(s) ds}}\vec{T}B^{\ast }(\xi _{n})\cdots B^{\ast }(\xi _{1})\,|\Omega \rangle \\
&=& \frac{1}{\prod_{k=1}^{n-1} \sqrt{\int_{0}^{\infty} |\xi_{k}(s)|^2 w_{k+1}(s) ds}}\\
&&\quad  \int_{\Delta _{n}}\xi _{n}(s_{n})\cdots \xi _{1}(s_{1})dB^{\ast
}(s_{n})\cdots dB^{\ast }(s_{1}) | \Omega \rangle,\\
&=&\frac{1}{\prod_{k=1}^{n-1} \sqrt{\int_{0}^{\infty} |\xi_{k}(s)|^2
w_{k+1}(s) ds}}\\
&&\quad \times \int_{0}^{\infty }\xi _{n}(s_{n})dB^{\ast
}(s_{n})\int_{0}^{s_{n}}\xi _{n-1}(s_{n-1})dB^{\ast }(s_{n-1}) \\
&&\cdots \int_{0}^{s_{2}}\xi _{1}(s_{1})dB^{\ast }(s_{1})| \Omega \rangle,
\end{eqnarray*}
where $\xi _{1},\ldots ,\xi _{n}$ are normalized wave packet shapes (i.e., $\int_{0}^{\infty} |\xi_k(s)|^2ds=1$) and the
integral is over the simplex 
\[
\Delta _{n}=\left\{ \left( s_{n},\cdots ,s_{1}\right) :s_{n}>s_{n-1}>\cdots
>s_{1}>0\right\} . 
\]
Note that we may define the time-ordering of a function $F_{n}$ of $n$ distinct time
arguments by 
\[
\vec{T}F_{n}(s_{n},\cdots ,s_{1}):=F_{n}\left( s_{\sigma (n)},\cdots
,s_{\sigma (1)}\right) 
\]
where $\sigma $ is the permutation such that $\left( s_{\sigma (n)},\cdots,s_{\sigma (1)}\right) \in \Delta _{n}$. In this case, 
\begin{eqnarray*}
|\overrightarrow{\xi _{n},\cdots ,\xi _{1}}\rangle &\equiv& \frac{1}{n! \prod_{k=1}^{n-1} \sqrt{\int_{0}^{\infty} |\xi_{k}(s)|^2
w_{k+1}(s) ds}}\\
&&\quad \int_{\lbrack
0,\infty )^{n}}\vec{T}\xi _{n}\otimes \cdots \otimes \xi _{1}(s_{n},\cdots
,s_{1})dB^{\ast }(s_{n})\cdots dB^{\ast }(s_{1})| \Omega \rangle 
\end{eqnarray*}

We note that when the wavepackets are all identical ($\xi_1 = \cdots \xi_n \equiv \xi $), the state reduces to 
the $n$-particle state 
\[
|\overrightarrow{\xi  ,\cdots ,\xi  }\rangle \equiv 
\frac{1}{\sqrt{n!}}\ B^{\ast }(\xi )^n |\Omega \rangle. 
\]

For these time-ordered multi-photon states, we take $D=n+1$, so that the auxiliary system will be realized
on $\mathfrak{h}_{\mathrm{aux}}=\mathbb{C}^{n+1}$ with basis $|0\rangle ,\ldots
,|n\rangle $. The initial state is taken to be the excited state $|n\rangle $%
. We take $S=I$,
\[
R(t)=\sum_{k=1}^{n}\lambda _{n+1-k}(t)|k-1\rangle \langle k|,
\]
and $H_{\mathrm{aux}}=0$, where 
\[
\lambda _{k}(t)=\frac{\xi _{k}(t)\sqrt{w_{k+1}(t)}}{\sqrt{\int_{0}^{\infty
}|\xi _{k}(s)|^{2}w_{k+1}(s)ds}\sqrt{w_{k}(t)}}, \]
and 
\[
w_{k}(t)=\frac{\int_{t}^{\infty }|\xi _{k}(s)|^{2}w_{k+1}(s)ds}{%
\int_{0}^{\infty }|\xi _{k}(s)|^{2}w_{k+1}(s)ds}, 
\]
for $k=1,2\ldots,n$, with 
$$
w_{n+1}(t)=1.
$$ 
This auxiliary system acts as the generator of a time-ordered $n$%
-photon state, as detailed in Appendix \ref{app:two-photon-generator} for time-ordered two-photon
states, and Appendix \ref{app:n-photon-generator} for the general time-ordered case with $n>2$.
The treatment of the two-photon generator in Appendix \ref{app:two-photon-generator}
provides a more detailed account of the underlying idea which is
subsequently generalized in Appendix \ref{app:n-photon-generator}. The CMP filter for
this time-ordered $n$-photon input state will be a $(n+1)\times (n+1)$ system of coupled
stochastic differential equations obtained from (\ref{eq:Homodyne_CMP_filter}%
) or (\ref{eq:number_CMP_filter}).

For filtering in  non-time-ordered multi-photon states of the form $B(\xi_1 )^\ast
 \cdots B(\xi_n)^\ast \, \vert \Omega \rangle$ see \cite{SZX13}. This, however,
requires a non-Markovian embedding approach.

\section{Conclusion}

\label{sec:conclusion} In this paper we have introduced a new class of
continuous matrix product states, which includes (continuous-mode) single
photon and time-ordered multi-photon states as special cases. We then derive the quantum
master equation and quantum filtering (quantum trajectory) equations for
Markovian open quantum systems driven by boson fields in the new class of
continuous matrix product states, that naturally take the form of
matrix-valued equations with operator entries. A Markovian generator of
time-ordered continuous-mode multi-photon states was also obtained, thus allowing the
quantum master and filtering equations for systems driven by time-ordered multi-photon
states to be readily derived using the general formulas of this paper, in
particular generalizing the Markovian embedding approach in \cite{GJNC12}.

\section*{Acknowledgements.} 
This work was supported by the Australian Research Council Centre of
Excellence for Quantum Computation and Communication Technology (project
number CE110001027), the Australian Research Council Discovery Project
Programme, EPSRC through Research Project EP/H016708/1, and Air Force Office
of Scientific Research (grant number FA2386-12-1-4075).

\section*{Appendices}
\appendices

\section{Time-ordered continuous-mode two-photon state generator}

\label{app:two-photon-generator}

In this appendix we develop a Markovian generator model for time-ordered continuous-mode
two-photon states of the field, and show explicitly that these generators
indeed produce time-ordered two-photon states. The ideas in this appendix are extended to
the general time-ordered multi-photon case in \ref{app:n-photon-generator}.

We consider an open three level system with levels $|0 \rangle=(0,0,1)^T$, $%
|1 \rangle=(0,1,0)^T$, $|2 \rangle=(1,0,0)^T$ coupled to a vacuum continuum
boson field via the (time-varying) coupling operator $L(t)= \lambda_2(t) |0
\rangle \langle 1| + \lambda_1(t) |1 \rangle \langle 2 |$, for some given
functions $\lambda_1(t)$ and $\lambda_2(t)$ that will be specified shortly.
We can thus write $L(t)$ as the $3 \times 3$ matrix-valued function 
\[
L(t)=\left[ 
\begin{array}{ccc}
0 & 0 & 0 \\ 
\lambda_1(t) & 0 & 0 \\ 
0 & \lambda_2(t) & 0
\end{array}
\right]. 
\]

For given wave packet shapes $\xi _{1}(t)$ and $\xi _{2}(t)$ (they need not
be the same shape), define $w_{2}(t)=\int_{t}^{\infty }|\xi _{2}(s)|^{2}ds$
and 
\[
w_{1}(t)=\frac{\int_{t}^{\infty }|\xi _{1}(s)|^{2}w_{2}(s)ds}{%
\int_{0}^{\infty }|\xi _{1}(s)|^{2}w_{2}(s)ds}, 
\]
and note that $w_{k}(0)=1$ and $w_{k}(\infty )=0$ for all $k$. From these
definitions then define $\lambda _{1}(t)$ and $\lambda _{2}(t)$ as 
\begin{eqnarray*}
\lambda _{2}(t) &=&\frac{\xi _{2}(t)}{\sqrt{w_{2}(t)}} \\
\lambda _{1}(t) &=&\frac{\xi _{1}(t)\sqrt{w_{2}(t)}}{\sqrt{\int_{0}^{\infty
}|\xi _{1}(s)|^{2}w_{2}(s)ds}\sqrt{w_{1}(t)}}.
\end{eqnarray*}
Before proceeding further, we note that 
\[
\exp \left( -\frac{1}{2}\int_{0}^{t}|\lambda _{j}(s)|^{2}ds\right) =\sqrt{%
w_{j}(t)} 
\]
for $j=1,2$. Let $|\psi (t)\rangle =(\psi _{2}(t),\psi _{1}(t),\psi
_{0}(t))^{T}\otimes |\Omega _{\lbrack t}\rangle $ ($|\Omega _{\lbrack
t}\rangle $ denotes the portion of the Fock vacuum on $[t,\infty )$)) be a
state vector process solving the QSDE 
\[
d|\psi (t)\rangle =\left( -\frac{1}{2}L^{\ast }Ldt+LdB^{\ast }(t)-L^{\ast
}dB(t)\right) |\psi (t)\rangle . 
\]
with initial condition $|\psi (0)\rangle =|2\rangle \otimes |\Omega \rangle $, 
where $|\Omega \rangle $ denotes the Fock vacuum. Since $|\psi (t)\rangle $
has a tensor product form with a vacuum component on the portion of the Fock
space from $t$ onwards, the QSDE can be simplified as 
\begin{eqnarray*}
d|\psi (t)\rangle &=&\left( -\frac{1}{2}L^{\ast }Ldt+LdB^{\ast }(t)\right)
|\psi (t)\rangle . \\
&=&\left[ 
\begin{array}{ccc}
-\frac{1}{2}|\lambda _{1}(t)|^{2}dt & 0 & 0 \\ 
\lambda _{1}(t)dB^{\ast }(t) & -\frac{1}{2}|\lambda _{2}(t)|^{2}dt & 0 \\ 
0 & \lambda _{2}(t)dB^{\ast }(t) & 0
\end{array}
\right] |\psi (t)\rangle .
\end{eqnarray*}
This leads to the following set of coupled equations for the component $\psi
_{2}$, $\psi _{1}$ and $\psi _{0}$ of $\psi $: 
\[
d\left[ 
\begin{array}{c}
\psi _{2}(t) \\ 
\psi _{1}(t) \\ 
\psi _{0}(t)
\end{array}
\right] \otimes |\Omega _{\lbrack t}\rangle =\left[ 
\begin{array}{c}
-\frac{1}{2}|\lambda _{1}(t)|^{2}\psi _{2}(t)dt \\ 
-\frac{1}{2}|\lambda _{2}(t)|^{2}\psi _{1}(t)dt+\lambda _{1}(t)\psi
_{2}(t)dB^{\ast }(t) \\ 
\lambda _{2}(t)\psi _{1}(t)dB^{\ast }(t)
\end{array}
\right] |\Omega _{\lbrack t}\rangle 
\]
with initial conditions $\psi _{2}(0)=1$, $\psi _{1}(0)=0$, and $\psi
_{0}(0)=0$. Using the definitions and properties of $w_{1}$, $w_{2}$, $%
\lambda _{1}$, and $\lambda _{2}$, the special structure of the coupled
equations allow them to be solved explicitly giving the solutions: 
\begin{eqnarray*}
\psi _{2}(t) &=&\exp \left( -\frac{1}{2}\int_{0}^{t}|\lambda
_{1}(s)|^{2}ds\right) =\sqrt{w_{1}(t)}|\Omega _{t]}\rangle , \\
\psi _{1}(t) &=&\exp \left( -\frac{1}{2}\int_{0}^{t}|\lambda
_{2}(s)|^{2}ds\right) \\
&&\quad \times \int_{0}^{t}\lambda _{1}(s_{1})\exp \left( \frac{1}{2}%
\int_{0}^{s_{1}}(|\lambda _{2}(s_{2})|^{2}-|\lambda
_{1}(s_{2})|^{2})ds_{2}\right) dB^{\ast }(s_{1})|\Omega _{t]}\rangle , \\
&=&\frac{\sqrt{w_{2}(t)}}{\sqrt{\int_{0}^{\infty }|\xi _{1}(s)|^{2}w_{2}(s)ds%
}}\int_{0}^{t}\xi _{1}(s_{1})dB^{\ast }(s_{1})|\Omega _{t]}\rangle , \\
\psi _{0}(t) &=&\int_{0}^{t}\lambda _{2}(\tau )\psi _{1}(\tau )dB^{\ast
}(\tau )|\Omega _{t]}\rangle , \\
&=&\frac{1}{\sqrt{\int_{0}^{\infty }|\xi _{1}(s)|^{2}w_{2}(s)ds}}%
\int_{0}^{t}\lambda _{2}(\tau )\sqrt{w_{2}(\tau )}\int_{0}^{\tau }\xi
_{1}(s_{1})dB^{\ast }(s_{1})dB^{\ast }(\tau )|\Omega _{t]}\rangle , \\
&=&\frac{1}{\sqrt{\int_{0}^{\infty }|\xi _{1}(s)|^{2}w_{2}(s)ds}}%
\int_{0}^{t}\xi _{2}(\tau )dB^{\ast }(\tau )\int_{0}^{\tau }\xi
_{1}(s_{1})dB^{\ast }(s_{1})|\Omega _{t]}\rangle .
\end{eqnarray*}

Note that as $t\rightarrow \infty $, $\psi _{2}(t)\rightarrow 0$ and $\psi
_{1}(t)\rightarrow 0$ since $\mathop{\lim}_{k\rightarrow \infty }w_{k}(t)=0$
for $k=1,2$. That is, as $t\rightarrow \infty $ the atom decays to its
ground state and the output field of the system tends to the state $\psi
_{0}(\infty )$. More precisely,
\[
\psi _{0}(t)=\frac{1}{\sqrt{\int_{0}^{\infty }|\xi _{1}(s)|^{2}w_{2}(s)ds}}%
\int_{0}^{t}\int_{0}^{\tau }\xi _{2}(\tau )\xi _{1}(s_{1})dB^{\ast
}(s_{1})dB^{\ast }(\tau )|\Omega _{t]}\rangle. 
\]
which shows that $\psi _{0}(t)$ converges as $t\rightarrow \infty $ to the
time-ordered two-photon state 
\begin{eqnarray*}
\psi_{0}(\infty )&=&\frac{1}{\sqrt{\int_{0}^{\infty }|\xi _{1}(s)|^{2}w_{2}(s)ds}} \int_{0}^{\infty }\int_{0}^{\tau }
\xi_{2}(\tau )\xi _{1}(s_{1})dB^{\ast }(s_{1})dB^{\ast }(\tau )|\Omega \rangle \\
&=& |\overrightarrow{\xi _{2},\xi _{1}}\rangle. 
\end{eqnarray*}
Moreover, since $\psi _{0}(\infty )$ is a bona fide pure
state vector on the field, we also note that 
\[
\left\| \int_{0}^{\infty }\int_{0}^{\tau }\xi _{2}(\tau )\xi
_{1}(s_{1})dB^{\ast }(s_{1})dB^{\ast }(\tau )|\Omega \rangle \right\| =\sqrt{%
\int_{0}^{\infty }|\xi _{1}(s)|^{2}w_{2}(s)ds}. 
\]

If $\xi_1(t)=\xi_2(t)\equiv \xi(t)$ then $\xi_1(s_1)\xi_2(\tau) = \xi(s_1)\xi(\tau)$ becomes  symmetric with respect
to its arguments $s_1$ and $\tau$, and so we have the identity:
$$
\int_{0}^{\infty }\int_{0}^{\tau }
\xi(\tau )\xi(s_{1})dB^{\ast }(s_{1})dB^{\ast }(\tau )|\Omega \rangle = \frac{1}{2} \int_{0}^{\infty }\int_{0}^{\infty }
\xi(\tau )\xi(s_{1})dB^{\ast }(s_{1})dB^{\ast }(\tau )|\Omega \rangle. 
$$
Moreover, also we have that $\int_{0}^{\infty }|\xi _{1}(s)|^{2}w_{2}(s)ds =\frac{1}{2}$. It follows that for the special case  $\xi_1(t)=\xi_2(t)=\xi(t)$ that
\begin{eqnarray*}
\psi_{0}(\infty )&=&\sqrt{2} \frac{1}{2}\int_{0}^{\infty }\int_{0}^{\infty} \xi(\tau )\xi(s_{1})dB^{\ast }(s_{1})dB^{\ast }(\tau )|\Omega \rangle, \\
&=& \frac{1}{\sqrt{2}} \int_{0}^{\infty }\int_{0}^{\infty} \xi(\tau )\xi(s_{1})dB^{\ast }(s_{1})dB^{\ast }(\tau )|\Omega \rangle,\\
&=& \frac{1}{\sqrt{2}} B^*(\xi)^2 |\Omega \rangle. 
\end{eqnarray*}

\section{Time-ordered continuous-mode $n$-photon state generator with $n>2$}
\label{app:n-photon-generator}
In this appendix, we generalize the time-ordered two-photon generator model treated in 
\ref{app:two-photon-generator} to general time-ordered $n$-photon case with $n>2$.

Consider an open $n+1$ level system with levels $|0
\rangle=(0,0,\ldots,0,1)^T$, $|1 \rangle=(0,0,0,\ldots,1,0)^T$, $\ldots$, $%
|n \rangle=(1,0,0,\ldots,0)^T$, coupled to a vacuum continuum boson field
via the (time-varying) coupling operator $L(t)= \sum_{k=1}^{n}
\lambda_{n+1-k}(t) |k-1 \rangle \langle k|$, for some given functions $%
\lambda_1(t),\ldots, \lambda_n(t)$ that will be specified shortly. We can
thus write $L(t)$ as the $(n+1) \times (n+1)$ matrix-valued function 
\[
L(t)=\left[ 
\begin{array}{cc}
0_{1 \times n} & 0 \\ 
\mathrm{diag(\lambda_{1}(t),\lambda_{2}(t),\ldots,\lambda_n(t))} & 0_{n
\times 1}
\end{array}
\right], 
\]
where $\mathrm{diag}(a_1,a_2,\ldots,a_m)$ denotes a diagonal matrix with
diagonal entries $a_1,a_2,\ldots,a_m$ from top left to bottom right.

For given wave packet shapes $\xi_1(t), \xi_2(t),\ldots \xi_n(t)$ (not
necessarily identical), define $w_n(t)=\int_{t}^{\infty} |\xi_n(s)|^2ds$,
and 
\[
w_{k}(t)=\frac{\int_{t}^{\infty} |\xi_k(s)|^2 w_{k+1}(s) ds}{%
\int_{0}^{\infty} |\xi_k(s)|^2 w_{k+1}(s) ds}, 
\]
recursively for $k=n-1,n-2,\ldots,1$. Note, in particular, that $w_k(0)=1$ and 
$w_k(\infty)=0$ for all $k$. From these definitions then define $\lambda_n(t)= 
\frac{\xi_n(t)}{\sqrt{w_n(t)}}$, and 
\[
\lambda_k(t) = \frac{\xi_k(t) \sqrt{w_{k+1}(t)}}{\sqrt{\int_{0}^{\infty}
|\xi_k(s)|^2 w_{k+1}(s) ds}\sqrt{w_k(t)}}, 
\]
recursively for $k=1,2,\ldots,n-1$. As was the case for time-ordered two-photon fields, we
verify from the definitions that 
\begin{eqnarray*}
\exp\left(-\frac{1}{2}\int_0^t |\lambda_j(s)|^2 ds \right) &=& \sqrt{w_j(t)}
\end{eqnarray*}
for $j=1,\ldots,n$. Let $|\psi(t)\rangle =(\psi_2(t),\psi_1(t),\psi_0(t))^T
\otimes |\Omega_{[t} \rangle$ ($|\Omega_{[t}\rangle$ denotes the portion of
the Fock vacuum on $[t,\infty)$) be a state vector process solving the QSDE 
\begin{eqnarray*}
d|\psi(t) \rangle &=& \left(-\frac{1}{2}L^*L dt +L dB^*(t) - L^*
dB(t)\right) |\psi(t) \rangle.
\end{eqnarray*}
with initial condition $|\psi(0)\rangle = |n\rangle \otimes |\Omega \rangle$%
, where $|\Omega \rangle$ denotes the Fock vacuum. Since $|\psi(t) \rangle$
has a tensor product form with a vacuum component on the portion of the Fock
space from $t$ onwards, the QSDE can be simplified as 
\begin{eqnarray*}
d|\psi(t) \rangle &=&\left(-\frac{1}{2}L^*L dt +L dB^*(t)\right) |\psi(t)
\rangle. \\
&=& \left(-\frac{1}{2}\left[ 
\begin{array}{cc}
\mathrm{diag}(|\lambda_1(t)|^2,\ldots,|\lambda_{n-1}(t)|^2,|\lambda_{n}(t)|^2) & 0_{n \times 1} \\ 
0_{1 \times n} & 0
\end{array}
\right] dt \right. \\
&&\quad \left. +\left[ 
\begin{array}{cc}
0_{1 \times n} & 0 \\ 
\mathrm{diag(\lambda_1(t),\ldots,\lambda_{n-1}(t),\lambda_n(t))} & 0_{n
\times 1}
\end{array}
\right] dB^*(t)\right) |\psi(t) \rangle.
\end{eqnarray*}
This leads to the following set of coupled equations for the component $%
\psi_0,\psi_1,\ldots, \psi_n$ of $\psi$: 
\begin{eqnarray*}
d\left[ 
\begin{array}{c}
\psi_n(t) \\ 
\vdots \\ 
\psi_1(t) \\ 
\psi_0(t)
\end{array}
\right] \otimes |\Omega_{[t} \rangle &=& \left(-\frac{1}{2}\left[ 
\begin{array}{c}
|\lambda_1(t)|^2 \psi_n(t) \\ 
\vdots \\ 
|\lambda_n(t)|^2 \psi_1(t) \\ 
0
\end{array}
\right]dt \right. \\
&&\quad \left. +\left[ 
\begin{array}{c}
0 \\ 
\lambda_1(t)\psi_n(t) \\ 
\vdots \\ 
\lambda_n(t)\psi_1(t)
\end{array}
\right]dB^*(t)\right) |\Omega_{[t} \rangle
\end{eqnarray*}
with initial conditions $\psi_n(0)=1$, and $\psi_k(0)=0$ for $%
k=0,1,\ldots,n-1$. Using the definitions and properties of $w_1,\ldots,w_n$,
and $\lambda_1,\ldots,\lambda_n$, as with the two-photon case the special
structure of the coupled equations allow them to be solved explicitly,
giving the solutions: 
\begin{eqnarray*}
\psi_n(t) &=& \sqrt{w_1(t)} |\Omega_{t]} \rangle, \\
\psi_{n-1}(t) &=& \frac{\sqrt{w_{2}(t)}}{\sqrt{\int_{0}^{\infty}
|\xi_{1}(s)|^2 w_{2}(s) ds}} \int_{0}^t \xi_1(s_1) dB^*(s_1) |\Omega_{t]}
\rangle, \\
&\vdots& \\
\psi_1(t) &=& \frac{\sqrt{w_{n}(t)}}{\prod_{k=1}^{n-1} \sqrt{\int_{0}^{\infty}
|\xi_k(s)|^2 w_{k+1}(s) ds}}\int_{0}^t \int_{0}^{s_{n-1}} \ldots
\int_{0}^{s_{2}} \xi_{n-1}(s_{n-1}) \cdots \xi_1(s_1) \\
&&\quad \times dB^*(s_1) \ldots dB^*(s_{n-2}) dB^*(s_{n-1}) |\Omega_{t]}
\rangle, \\
\psi_0(t) &=& \frac{1}{\prod_{k=1}^{n-1} \sqrt{\int_{0}^{\infty} |\xi_k(s)|^2
w_{k+1}(s) ds}}\int_{0}^t \int_{0}^{s_n} \ldots \int_{0}^{s_2}
\xi_{n}(s_{n})\xi_{n-1}(s_{n-1}) \cdots \\
&& \quad \xi_{1}(s_1) dB^*(s_1) \cdots dB^*(s_{n-1}) dB^*(s_{n}) |\Omega_{t]}
\rangle.
\end{eqnarray*}

Note that as $t\rightarrow \infty $, $\psi _{k}(t)\rightarrow 0$ as $%
t\rightarrow \infty $ for $k=1,\ldots ,n$ since $\mathop{\lim}_{k\rightarrow
\infty }w_{k}(t)=0$. That is, as $t\rightarrow \infty $ the atom decays to
its ground state and the output field of the system tends to the state $\psi
_{0}(\infty )$.  Taking the limit, $\psi _{0}(t)$ converges as $t\rightarrow \infty $ to the time-ordered $n$-photon state 
\[
\psi _{0}(\infty )=|\overrightarrow{\xi _{n},\cdots ,\xi _{1}}\rangle . 
\]
Moreover, since $\psi _{0}(\infty )$ is a bona fide pure
state vector on the field, we note that 
\begin{eqnarray*}
&&\left\| \int_{0}^{\infty }\int_{0}^{s_{n}}\ldots \int_{0}^{s_{2}}\xi
_{n}(s_{n})\xi _{n-1}(s_{n-1})\cdots \xi _{1}(s_{1})dB^{\ast }(s_{1})\cdots
dB^{\ast }(s_{n-1})dB^{\ast }(s_{n})|\Omega \rangle \right\| \\
&&\qquad =\prod_{k=1}^{n-1}\sqrt{\int_{0}^{\infty }|\xi
_{k}(s)|^{2}w_{k+1}(s)ds}.
\end{eqnarray*}

In the special case where $\xi_1=\xi_2=\ldots=\xi_n\equiv \xi$ then
$\xi_n(s_n)\xi_{n-1}(s_{n-1}) \cdots \xi_1(s_1)=\xi(s_n)\xi(s_{n-1}) \cdots \xi(s_1)$
becomes symmetric with respect to the arguments $s_1,s_2,\ldots,s_{n}$ and we have
\begin{eqnarray*}
&&\lefteqn{\int_{0}^{\infty }\int_{0}^{s_{n}}\ldots \int_{0}^{s_{2}}\xi
(s_{n})\xi(s_{n-1})\cdots \xi(s_{1})dB^{\ast }(s_{1})\cdots
dB^{\ast }(s_{n-1})dB^{\ast }(s_{n})|\Omega \rangle} \\
&& =  \frac{1}{n!} \int_{0}^{\infty }\int_{0}^{\infty}\ldots \int_{0}^{\infty}
\xi(s_{n})\xi(s_{n-1})\cdots \xi(s_{1})dB^{\ast }(s_{1})\cdots
dB^{\ast }(s_{n-1})dB^{\ast }(s_{n})|\Omega \rangle
\end{eqnarray*}
and 
\begin{eqnarray*}
\prod_{k=1}^{n-1} \int_{0}^{\infty} |\xi_k(s)|^2
w_{k+1}(s) ds =\frac{1}{n!}. \\
\end{eqnarray*}
Therefore, in this case
\[
\psi _{0}(\infty )=\frac{1}{\sqrt{n!}}B^{\ast}(\xi)^n |\Omega \rangle. 
\]

\end{document}